\documentclass{svjour3}                     
\smartqed  
\usepackage{graphicx}

\begin{document}

\title{Sunspots: from small-scale inhomogeneities towards a global theory}

\titlerunning{Sunspot inhomogeneities}  

\author{Rolf Schlichenmaier
}


\institute{R. Schlichenmaier \at
              Kiepenheuer-Institut f\"ur Sonnenphysik, Sch\"oneckstr. 6, 79104 Freiburg, Germany \\
              Tel.: +49-761-3198212\\
              \email{schliche@kis.uni-freiburg.de}           
}

\date{Date: \today /  Received:  / Accepted: }

\maketitle

\begin{abstract}
The penumbra of a sunspot is a fascinating phenomenon featuring complex velocity and magnetic fields. It challenges both our understanding of radiative magneto-convection and our means to measure and derive the actual geometry of the magnetic and velocity fields. In this contribution we attempt to summarize the present state-of-the-art from an observational and a theoretical perspective.

We describe spectro-polarimetric measurements which reveal that the penumbra is inhomogeneous, changing the modulus and the direction of the velocity, and the strength and the inclination of the magnetic field with depth, i.e., along the line-of-sight,  and on spatial scales below 0.5 arcsec. Yet, many details of the small-scale geometry of the fields are still unclear such that the small scale inhomogeneities await a consistent explanation.

A simple model which relies on magnetic flux tubes evolving in a penumbral "background" reproduces some properties of sunspot inhomogeneities, like its filamentation, its strong (Evershed-) outflows, and its uncombed geometry, but it encounters some problems in explaining the penumbral heat transport. Another model approach, which can explain the heat transport and long bright filaments, but fails to explain the Evershed flow, relies on elongated convective cells, either field-free as in the gappy penumbra or filled with horizontal magnetic field as in Danielson's convective rolls. Such simplified models fail to give a consistent picture of all observational aspects, and it is clear that we need a more sophisticated description of the penumbra, that must result from simulations of radiative magneto-convection in inclined magnetic fields. First results of such simulations are discussed. The understanding of the small-scales will then be the key to understand the global structure and the large-scale stability of sunspots. 

\keywords{First keyword \and Second keyword \and More}
\end{abstract}

\section{Introduction}
\label{intro}

Magnetic fields on the Sun exist in a large variety of phenomena and interact in various ways with the plasma and the radiation. In the convection zone large and small scale magnetic fields are generated. These magnetic fields are partially transported into the outer layers of the Sun, i.e., into the chromosphere and the corona. The most prominent example of a magnetic phenomenon is a sunspot as seen in the photosphere. A typical sunspot has a lifetime of a few weeks and has a size of about 30 granules. The magnetic field strength spans from 1000 to 3000 Gauss in the deep photosphere, summing up to a magnetic flux of some $10^{22}$ Mx.

The magnetic field of a sunspot extends into the interior as well as into the outer layers of the Sun. The most detailed information of sunspots is obtained in the photosphere. The topology of the magnetic field above and beneath the photosphere is poorly understood. In particular our knowledge of the magnetic field extension into the interior presents a theoretical challenge. Direct measurements of the sub-photospheric structure are impossible, but at least for the larger scales, indirect methods are being explored in the framework of local helioseismology (cf. Gizon, these proceedings).

\paragraph{Time scales:}

Although the sunspot is a coherent phenomenon on large spatial and temporal scales, it seems crucial to realize that it is not static, but finds a dynamical equilibrium: A variety of small-scale features evolve on a dynamic time scale to produce a large-scale coherent structure on long time scales. This ''fine structure'' is complex and is seen in white light images in form of umbral dots, light bridges, bright and dark penumbral filaments, penumbral grains, dark-cored bright filaments, penumbral twists, and other features. This intensity fine structure corresponds to a fine structure of the velocity field and the magnetic field, which will be described below. The dynamic fine structure forms a globally stable sunspot and it is the goal of sunspot physics to understand how an ensemble of short-lived features with small scales is organized to form a coherent large and long-living sunspot.

\section{Energy transport in umbra and penumbra}\label{sec:1}

The coolness of sunspots relative to the surrounding quiet Sun is readily explained by the tension of the magnetic field which tends to suppress convective motions. It is more difficult to understand why sunspots are as hot as they are: Neither radiative transport nor heat conduction can account for the surface brightness of sunspots. Hence convection cannot be fully suppressed and the energy must be transported by convective flows. Indeed, the fine structure manifests the inhomogeneities of the magnetic and velocity field and testifies that the energy transport in sunspots happens on small spatial scales by the motion of plasma. Yet, the crucial question is about the interaction between convective flows, the magnetic field, and the radiation. Are the flows non-magnetic or magnetic? What is their intrinsic spatial scale? Do coherent up- and downflows exist, similar to the granulation in the quiet Sun? 

\paragraph{Jelly fish and field-free gaps:}
Parker (1979) has introduced the jelly fish model in which the sub-photospheric magnetic field separates into individual bundles of field lines, resulting in gaps free of magnetic field. The gaps between these bundles open up into very deep layers, being connected to the quiet Sun convection. Within these cracks, the field-free plasma would convect and transport heat upwards. An umbral dot would correspond to the peak of a field-free gap. More recently, Spruit \& Scharmer (2006) suggested that such field-free gaps in the inclined magnetic field of the penumbra may result in elongated bright filaments, instead of in point-like dots, thereby proposing an explanation for the brightness of the penumbra. The surplus brightness of the penumbra relative to the umbra would then be due to the fact that the convective cell can become larger in the more inclined and weaker magnetic field as in the less inclined (more vertical) and stronger field of the umbra. 

\paragraph{Stability of sunspots and monolithic models:} Sunspots are stable relative to the dynamical time, i.e., Alfv{\'e}n waves are estimated to travel across a spot in about 1h, while the life time is in order of weeks. How can it be that all this dynamic fine structure constitutes a spot which is stable? The question of stability can be addressed if one assumes a "simple"  vertical magnetohydrostatic magnetic flux tube that fans out with heigth. In such models the heat transport is attributed to (magneto-) convection, but is parametrized by a reduced mixing length parameter (Jahn 1989, Deinzer 1965). The dynamic fine structure is ignored and only their averaged effect on the stratification for umbra and penumbra is accounted for. The configuration is in magneto-static equilibrium together with a hydrostatic equilibrium vertically and with a total pressure balance between the umbra, penumbra, and quiet Sun horizontally (see e.g. Jahn \& Schmidt 1994, Pizzo 1990). This configuration can be stable against the interchange instability (Meyer et al. 1977), at least in the first 5 Mm or so beneath the photosphere (Jahn 1997). In these upper layers of the convection zone the inclination of the interface between spot and surrounding is so large that buoyancy forces make the spot to float on the granulation. In deeper layers, beyond 5 Mm, the inclination of the outermost magnetic field line, i.e., the magnetopause, is small relative to the vertical. There, interchange (fluting) instability is no longer suppressed by buoyancy effects, and the magnetic configuration of a monolithic sunspot is unstable. Indeed, it has been proposed that the magnetic field strength progressively weakens in these deep layers shortly after the formation of a sunspot. The decreasing field strength, the convective motions, and the interchange instability dynamically disrupt the sunspot magnetic field from the deeper roots (Sch\"ussler \& Rempel 2005). Hence, the magnetic field in the deeper layers may be dispersed, but the floating part of the sunspot is stable.

\section{Inhomogeneities in umbra and penumbra}

For an extensive review of the sunspot structure, we refer the reader to an instructive overview by Solanki (2003).
 
\subsection{Umbral dots}

The umbra of a sunspot harbors dynamic inhomogeneities. They are observed as dot-like bright spots with typical sizes of half an arcsec or less, embedded in a more uniform and darker background. These umbral dots seem to be present in all sunspots, although their intensity varies a lot. In some spots they can be almost as bright as bright penumbral filaments, in other spots their intensity is much smaller. In the latter case, the dot-like intensity variations occurs in a background that also shows a lower intensity.

\paragraph{The physics of umbral dots:} Umbral dots are an obvious signature of convection, yet it is not so obvious to understand the type of convection that leads to umbral dots. In the field-free gap idea of Parker, the convection is confined by the strong surrounding magnetic field, such that the column of convection narrows upwards and only a small brightening is seen at the surface. Observationally, it is established that the magnetic field in umbral dots is weaker than in the surroundings and that an upflow of at least a few hundred m/s is associated with them (Socas-Navarro et al. 2004, Rimmele 2004, 2008, Bharti et al. 2007). The latter two observations also establish the presence of dark lanes across umbral dots.

The most recent simulations of radiatively driven magneto-convection in strong vertical magnetic field (Sch\"ussler \& V\"ogler 2006) result in local convective cells which produce umbral dots as well as their dark lanes. These cells barely touch the photosphere, similar as in Parker's idea. The cells extend downward for a few Mm, in the first 1 Mm the cells have a weak magnetic field strength. The weak field strength is caused by magnetic flux expulsion, i.e. convection advects the magnetic field (as in Weiss 1964). In the simulations the magnetic field strength in the cells amounts to a few hundred Gauss, but this number may decrease if magnetic dissipation is reduced in more advanced simulation runs. In any case, in deeper layers the magnetic field strength increases considerably. Hence the cells do not connect to field-free plasma in deeper layers. In this respect these new simulations change our model vision of umbral dots. Now, we may conceive that the umbra is an overall monolithic fully magnetic structure, in which the fine structure is a local disturbance. The dots are produced locally by magneto-convection processes, which are needed for the energy transport.

\subsection{Penumbral inhomogeneities}

\subsubsection{Morphological description}
The penumbra is a manifestation of small-scale structure. The variety of the penumbral intensity fine structure is described in detail in the contribution to this volume by G\"oran Scharmer. In essence, there are bright and dark filaments, as well as penumbral grains.  It turns out that bright filaments have dark cores  (Scharmer et al. 2002, S\"utterlin et al. 2004) and that intensity twists exist along bright filaments (Ichimoto et al. 2007b) on spatial scales of about 0.2 arcsec.
The challenge consists in measuring the spectroscopic and spectropolarimetric signatures of this fine structure in order to derive their thermodynamic properties as well as their velocity and magnetic field. Only recently, with the technological advance of adaptive optics and with observations from space, it has become possible to acquire such high spatial resolution data for exposure times as long as 5 sec or more. This is a necessity to collect enough photons to have high spatial, spectral, and polarimetric resolution.

At a spatial resolution of better than half an arcsec, it can be demonstrated that not only the intensity and velocity, but also the magnetic field concists of a filamentary structure (Title et al. 1993, Langhans et al. 2005, Tritschler et al. 2007, Ichimoto et al. 2007a, 2008). Actually, at a spatial resolution of better than half an arcsec, all physical quantities in the penumbra show small-scale variations and predominantly filamentary (radially elongated) features. 

However, the penumbra looks fairly uniform at a spatial resolution worse than 1 arc sec. At  this lower spatial resolution, i.e. in average, the penumbra is brighter than the umbra, but less bright than in the surrounding granulation. But even if the penumbra is less bright in average, the small scale peak-to-peak intensity variation in the penumbra is larger than in the granulation, and the spatial scales of the variations are smaller than in the granulation. The same is true for velocities in the penumbra. Line-of-sight velocities in the penumbra of more than 5\,km/s have been derived from Doppler shifts of photospheric lines (e.g., Wiehr 1995) and radial flow channels with widths of less than half an arcsec are observed (e.g., Tritschler et al. 2004, Rimmele \& Marino 2006). 

\begin{figure}
\resizebox{10.1cm}{!}{\includegraphics[bb=0 99 2125 1417]{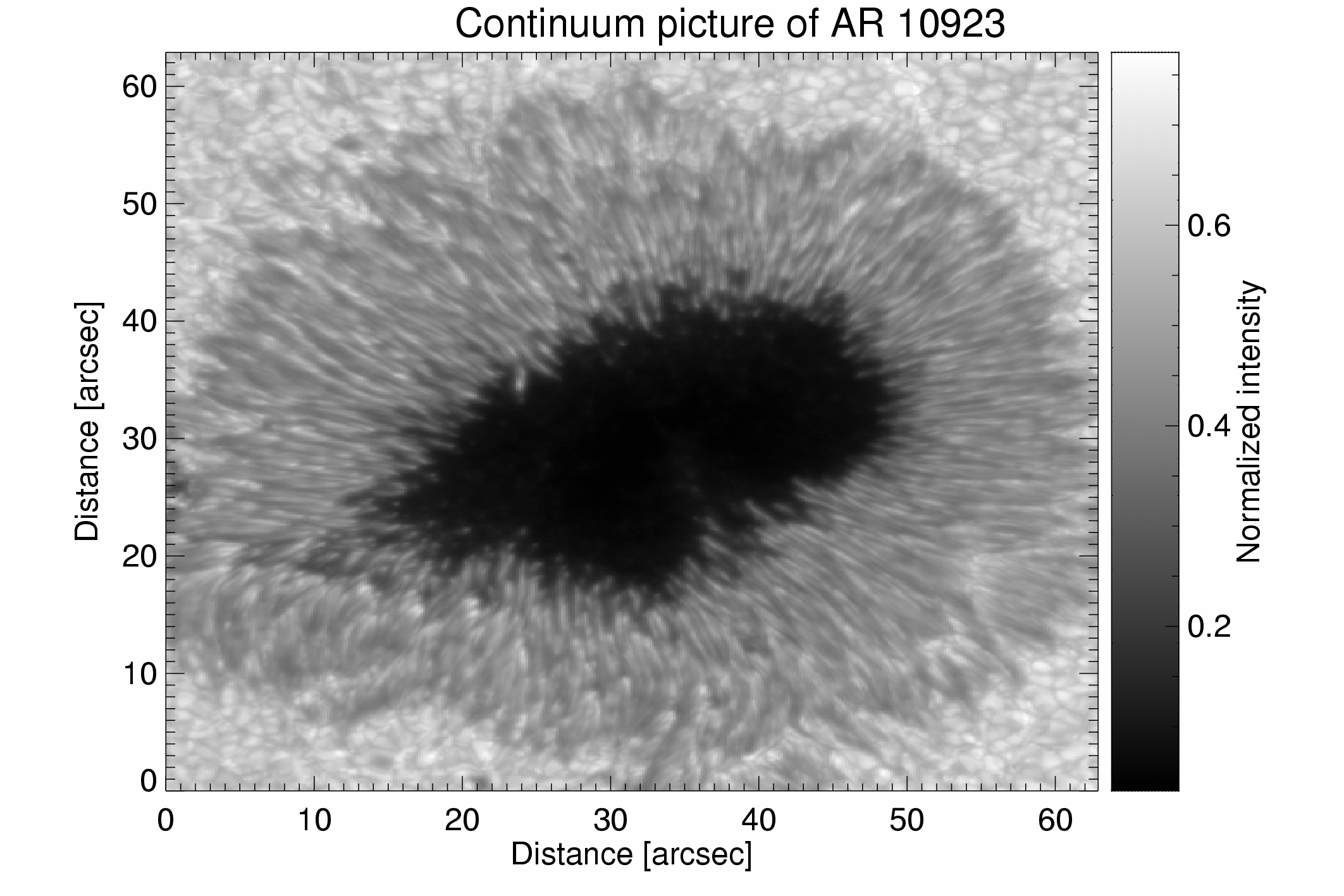}}\\
\resizebox{10.1cm}{!}{\includegraphics[bb=0 95 2125 1417]{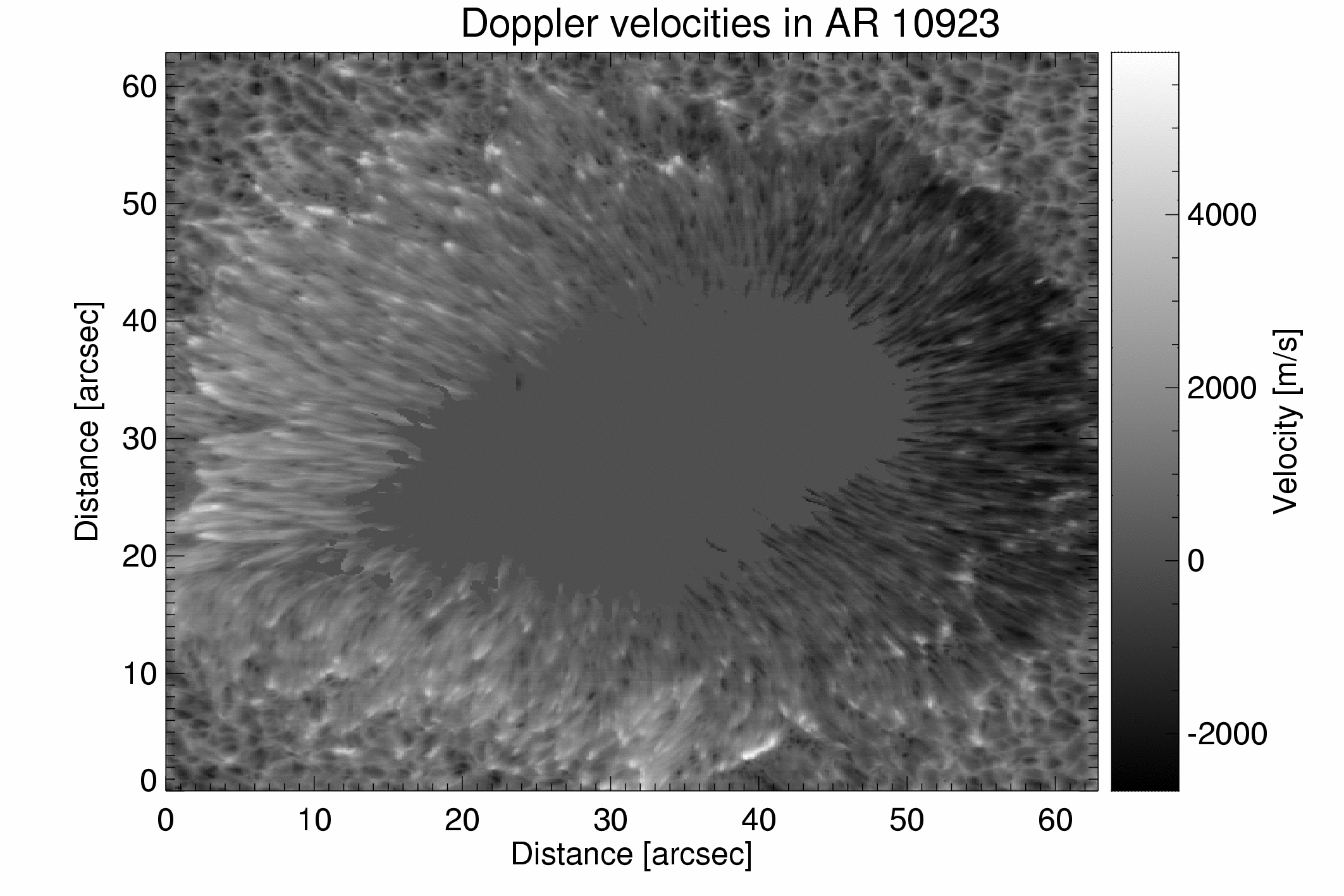}}\\
\resizebox{10.1cm}{!}{\includegraphics[bb=0 95 2125 1417]{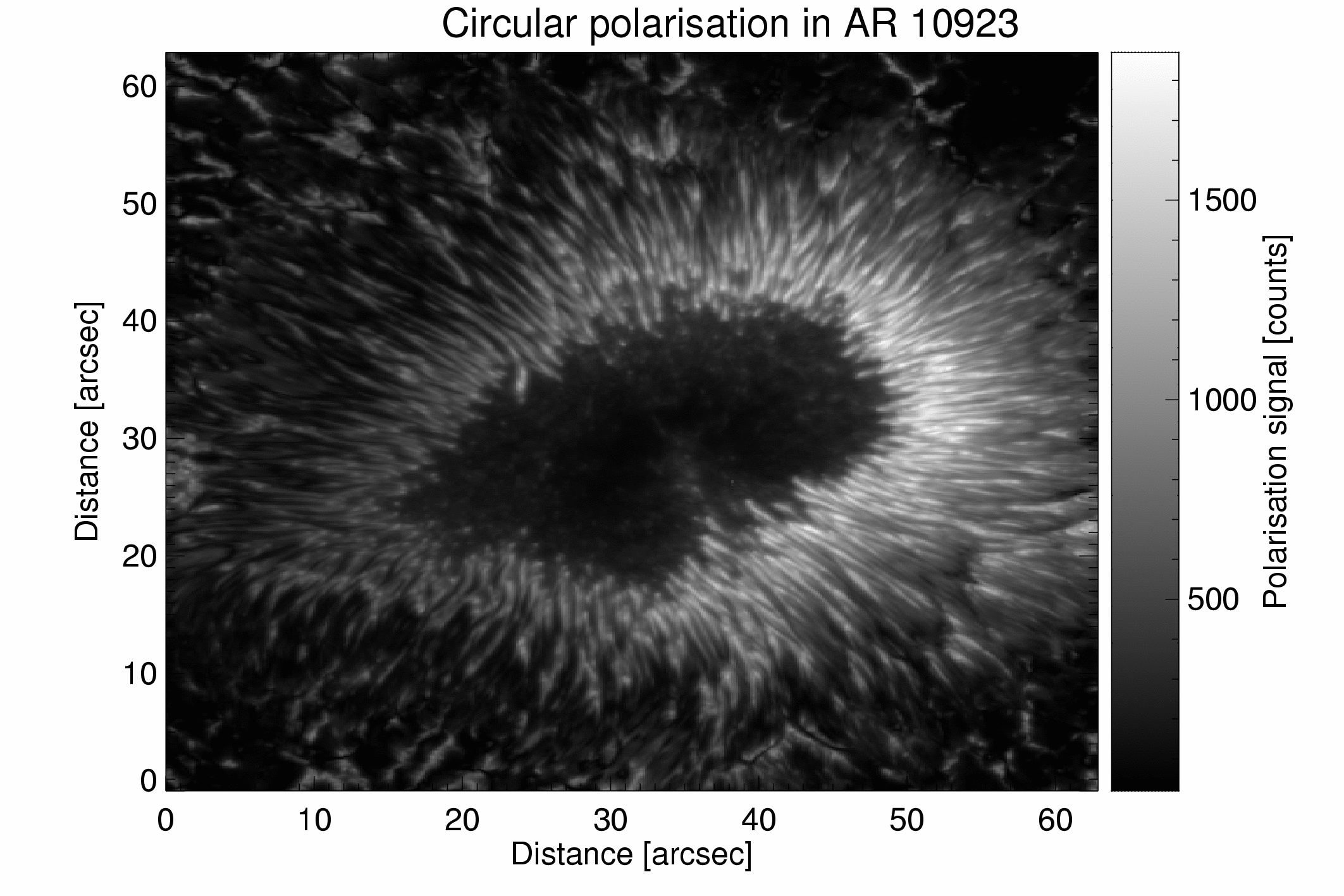}}\\
\caption{Maps of intensity, LOS velocity, and  circular polarization of sunspot (12 Nov 2006, $\theta=30^{\circ}$) from Fe I 630.2 nm taken with the spectropolarimeter SP attached to the SOT onboard Hinode.} 
\end{figure}

\subsubsection{Evershed flow, uncombed magnetic field,  and NCP}
For understanding the nature of the penumbral fine structure, it is essential to know the topology of the velocity field and the magnetic field. The first attempt to measure the flow field was undertaken by Evershed in 1908 (see Evershed, 1909) in order to test Hale's tornado theory of sunspots. Yet, instead of a circular flow, Evershed found a radial outflow of plasma, and until today we lack a consistent theory for  sunspots. Before we discuss the progress in modeling the characteristic feature of the penumbra, we discuss important observational aspects.

\paragraph{The flow field:} With high spatial resolution, it is now established that the flow has a filamentary structure (Tritschler et al. 2004, Rimmele \& Marino 2006). On average, the flow has a small upward component in the inner penumbra and a small  downward component in the outer penumbra (Schlichenmaier \& Schmidt 2000, Schmidt \& Schlichenmaier 2000, Tritschler et al. 2004, Langhans et al. 2005). Recent observations have revealed that radially aligned up- and downflows exist on small scales next to each other (Sainz Dalda \& Bellot Rubio 2008). Regarding the photospheric height at which the flow exists, there is convincing evidence that the flow is predominantly present in the very deep photosphere, i.e., beneath $\tau=0.1$ (Maltby 1964, Schlichenmaier et al. 2004, Bellot Rubio et al. 2006). The flow velocities measured in the penumbra are substantially larger than what is measured in the granulation. Individual penumbral profiles exhibit line satellites that are Doppler shifted by up to 8 km/s (e.g., Wiehr 1995). From inversions, velocities well above 10 km/s have been found by del Toro Iniesta et al. (2001). Bellot Rubio et al. (2004) find an azimuthally averaged Evershed flow velocity of about 6.5 km/s, with local peaks of more than 10 km/s, based on two component inversions (see below). The small-scale flow field of dark cored bright filaments is discussed in the context of convective roll models (at the end of Sect.~\ref{sec.convective}).

\paragraph{The magnetic field:} Attempts to describe the magnetic field as being uniform along the line of sight are clearly inconsistent with the measured Stokes $Q(\lambda)$, $U(\lambda)$, and $V(\lambda)$ profiles (e.g., Westendorp Plaza et al. 2001a, 2001b). In particular, the penumbral V-profiles with 3 or more lobes cannot be explained by one component, even if unresolved Doppler-shifted components are assumed (Schlichenmaier \& Collados 2002).
Therefore, it was proposed that the magnetic field is interlocked or in other words uncombed (Solanki \& Montavon 1993). In order to keep things as simple as possible, the magnetic field is assumed to have two components with different directions.  Indeed, if the observed Stokes profiles with a spatial resolution of about 1 arcsec are interpreted with two components by means of inversions techniques, the fit to the observations is much better than with only one component (Bellot Rubio 2004, Bellot Rubio et al. 2003, 2004, Borrero et al. 2004, 2005, Beck 2008). Such inversions yield one less inclined magnetic component that is only slightly Doppler shifted, and a second magnetic field component that is somewhat weaker and more inclined, i.e., approximately horizontal. This second component carries the Evershed flow, with spatially averaged flow speeds of about 6.5 km/s.

These inversions also show that the magnetic field of the second component is aligned with the associated flow, pointing slightly upwards in the inner and slightly downward in the outer penumbra. The inclination of the first magnetic field component increases from some 30 degree at the umbral-penumbral boundary to some 60 degrees at the outer penumbral boundary. Inversions which are optimized to locate the width and height of the flow layer find that the flow is present in the very deep atmosphere, in the continuum forming layers (Bellot Rubio 2003, Borrero et al. 2006, Jurcak et al. 2007, Jurcak \& Bellot Rubio 2008). Indeed, the width can hardly be determined, since the lower end of the flow layer is found to be beneath $\tau=1$. 

At 0.3 arcsec spatial resolution, spectropolarimetric measurements reveal that, at least in the inner penumbra, the more inclined magnetic component which carries the flow is associated with the dark cored bright filaments. Individual dark cores have a smaller degree of circular polarization than their lateral brightenings (Langhans et al. 2007). A thorough analysis shows that the latter statement is also true for the total polarization and that the dark core magnetic field is weaker and more inclined than in the lateral brightenings (Bellot Rubio et al. 2007). Additionally, these studies confirm that the dark cores harbor strong Evershed flows.

\paragraph{The magnetic canopy:} Outside the white-light boundary of the penumbra, the inclined magnetic field continues into the chromosphere, forming a magnetic chromospheric canopy in the surroundings of the sunspot, rising with distance to the spot up to a height of approximately 800 km (Solanki et al. 1992). In the canopy a radial outflow is present which is interpreted as the continuation of the Evershed flow (Solanki et al. 1992, Rezaei et al. 2006). However, it is estimated that only a few tenth of the flow mass is seen in the canopy. The rest of the penumbral Evershed flow must disappear within the penumbral downflow regions.

\paragraph{The net circular polarization (NCP):} The NCP, $\int V(\lambda)\, {\rm d} \lambda$, is a quantity that intimately links the flow and the magnetic field: NCP can only be non-zero, if and only if velocity gradients along the line of sight are present (e.g., Sanchez Almeida \& Lites 1992). The magnitude and the size of the NCP depends on the gradient of the line of sight velocity, but also on the gradients in the magnetic field strength, inclination, and azimuth (Landolfi \& Landi degl'Innocenti 1996, M\"uller et al. 2002,  2006, Borrero et al. 2008). A predominantly horizontal flow channel embedded in a less inclined background magnetic field successfully explains some symmetry properties of NCP maps of sunspots (Schlichenmaier et al. 2002) as well as some properties of the center to limb variation of NCP (Martinez Pillet 2000, Borrero et al. 2007).

Yet, some recent interpretations of NCP maps require that the flow component should be associated with stronger magnetic field (Tritschler et al. 2007, Ichimoto et al. 2008), rather than being associated with the same or weaker magnetic field in the flow channels, as we would expect from the models. Since there are also other indications for these stronger magnetic fields (e.g., Bellot Rubio 2003, Cabrera Solana et al. 2008, Borrero \& Solanki 2008), the concept of embedded flow channels will need to be reviewed taking into account these new measurements.

\paragraph{Magnetized or non-magnetized flow:} In terms of modeling the Evershed flow, it is crucial to know whether or not the flow is magnetized. While NCP can be generated by a field-free flow in a magnetized environment (e.g., Steiner 2000), the observed $V$ profiles in certain locations in the penumbra show more than two lobes (e.g., Schlichenmaier \& Collados 2002, Beck 2008). These additional lobes must be generated by an additional magnetic component: A non-magnetic component may produce a line asymmetry of Stokes-$I$ and a non-zero NCP, but it cannot produce additional lobes in Stokes-$V$. For this it needs to be magnetized! And after all, the inversion results based on two components (see above)  demonstrate that the Doppler-shifted "second" component is magnetized. Hence, we are convinced that any model for the penumbra needs to account for a magnetized Evershed flow.

\section{Penumbral Models}

The previous section stresses the point that the penumbra is a phenomenon of complex interaction of magneto-convective forces and radiation in a regime of inclined magnetic field of intermediate strength. One simplified view on this problem is to consider a separation between convective plumes and a magnetic configuration as it is done in the field-free gap model. Another simplified view is by dealing with the problem in ideal MHD, in which the thin flux tube approximation is applicable. The latter perspective is taken in the siphon flow model and the dynamic extension of it, the moving tube model. Yet, for a full understanding it seems necessary to take into account dissipative magneto-convection driven by radiation. But we want to stress that simplified models often help to isolate the dominating physical processes, and to understand the essentials.

\subsection{Convective models}\label{sec.convective}

Originally proposed by Parker to explain the umbral dots, Spruit \& Scharmer (2006) and Scharmer \& Spruit (2006) extended the concept of the field-free gaps to explain the bright penumbral filaments and they realized that such a configuration may also produce the dark cores within bright filaments, caused by a subtle radiative effect at the top of the field-free gap. The idea of field-free gaps in the penumbra is that the inclined penumbral magnetic field produces bright elongations instead of dots. The gaps are supposed to be void of magnetic field and to be connected to the surrounding quiet Sun. Within the gaps, overturning convection transports ample amounts of heat which would account for the brightness of the penumbra. The convective flow field is directed upwards along the central lane of the filament and downward at the edges of the long sides of the filaments. Within the field-free gap there may exist a radial outflow that corresponds to the Evershed flow. 
The problem with this description is that the Evershed flow which is observed to be magnetized need to be non-magnetized in the field-free gap.

The field-free gap model is in many respects similar to the model of the convective rolls proposed in 1961 by Danielson (see also Grosser 1991 for a numerical investigation on this model). Convective rolls lie radially aligned next to each other. Two such rolls would form one filament as they rotate in opposite direction, producing an upflow in the central lane and a downflow at the lateral lanes. Danielson assumed that a horizontal magnetic field component would be associated with the rolls. This model has been discarded for two reasons: (1) There was no evidence for the corresponding convective flow field, and (2) a major fraction of the magnetic flux in the penumbra is directed upwards, and not horizontal. However, reason (1) depends on spatial resolution and the issue is not settled yet, as we cannot rule out the existence small amplitude vertical motions of a few hundred m/s. Reason (2) could be overcome by assuming that the rolls are separated by less inclined (more vertical) magnetic field lines, which constitute a more or less static background magnetic field. And, magnetized rolls interlaced by a static background field that is less inclined relative to the vertical would also meet the observational requirements of two magnetic components in the penumbra. In this respect, at least in principle, it is possible that the horizontal magnetic component carries an Evershed flow.

The problem here is that -- up to now -- there is only little support for downflows along the edges of bright filaments, although there is some indication for weak upflows within the dark cores (e.g., Zakharov et al. 2008). Rimmele (2008) does find a convective-roll like flow field in a filament that extends into the umbra for a sunspot close to disk center, while Bellot Rubio et al. (2005) did not find indications for up and down flows associated with a dark-cored bright filament at disk center. Zakharov et al. (2008) observe a very small downward velocity component. The latter authors argue that the downflow may be obscured by the upflows, and that convective rolls exist. Hence, the crucial question of vertical flows in the penumbra needs to be reconsidered. In order to minimize the effects of the horizontal radial outflow and of possible flows in azimuthal direction, sunspot observations at disk center are needed to learn about the presumably small vertical flow component.

\subsection{Ideal magneto-convection}

Another simplified view of the problem is by restricting oneself to ideal MHD. In the self-consistent magneto-static tripartite sunspot model of Jahn \& Schmidt (1994) the surplus brightness of the penumbra relative to the umbra is produced by a heat transfer through the magnetopause, i.e. through the interface between the quiet Sun and the penumbra. This additional heat is thought to be distributed horizontally by interchange convection of magnetic flux tubes. The idea of dynamic magnetic flux tubes is compatible with the observationally finding of multiple magnetic components in the penumbra. 

This motivated the study of the dynamics of a single thin magnetic flux tube as it evolves in a 2D static model background (Schlichenmaier et al. 1998a,b). However, these studies did not confirm the concept of interchanging magnetic flux tubes which distribute heat horizontally. Instead, these studies created a new picture:
The simulated tube lies along the magnetopause of the tripartite sunspot model and is taken to be a bundle of magnetic field lines with penumbral properties. Initially the tube is in magneto-static equilibrium. However, at a magnetopause that is sufficiently inclined, radiative heat exchange between the tube and the hotter quiet Sun triggers an instability:
A thin magnetic flux tube that initially lies along the magnetopause, (a) feels the hotter quiet Sun, (b) heats up by radiation most effectively just beneath the photosphere, (c) expands, (d) rises through the subphotospheric convectively unstable stratification, and (e) develops an upflow along the tube, which brings hot subphotospheric plasma into the photosphere. (f) This hot upflow cools radiatively in the photosphere and streams radially outwards with supercritical velocity. The radiative cooling sustains the gas pressure gradient that drives the flow. (g) The ouflow intrudes the convectively stable photosphere up to a height of some 50 to 100 km. The equilibrium height is determined by the balance of the diamagnetic force which pulls the conducting tube upwards toward decreasing magnetic field strength and the downward acting buoyancy which increases as the tube is being pulled up in a convectively stable stratification.

\paragraph{Weak magnetic field at footpoint:} The gas pressure gradient that drives the flow is caused by a surplus gas pressure building up inside the part of the tube that rises through the subphotospheric stratification. At the foot-point, i.e., the intersection of the tube with the transition layer from convectively unstable to stable, the gas pressure is high, and in order to balance the total pressure with the surroundings, the magnetic field strength is strongly decreased relative to the surroundings. In this sense the upflow foot-points can be considered as regions of weak magnetic field strength. In other words, the moving tube model is a magneto-convective mode which consists of a region of weak-field plasma that harbors hot up-flows and that travels inwards.

In principle the effect leading to the up and out flow works like an inverse convective collapse: In the classical convective collapse the plasma in the tube is cooled and a downflow occurs. Here, the heating of the plasma results in an upflow, and consequently the magnetic field strength in the tube decreases as the flow continues. In the photosphere, the gas pressure gradient is sustained by radiative cooling.

The moving tube scenario successfully explains a number of observational findings: (i) Penumbral grains are the photospheric footpoints of the tube, where the hot and bright plasma enters the photosphere. (ii) The upflow turns horizontally outwards in the photosphere and cools radiatively until it reaches temperature equilibrium. This determines the length of the penumbral grains. (iii) The footpoints migrate inward, as many observed penumbral grains do (e.g., Sobotka \& S\"utterlin 2001). (iv) The horizontal outflow corresponds to the Evershed flow. (v) The tube constitutes a flow channel being embedded in a background magnetic field. This is in agreement with the uncombed penumbra, and produces realistic maps of NCP.

\paragraph{Magneto-convective overshoot:} An interesting effect that can be studied with the idealized moving tube model, is related to overshooting (Schlichenmaier 2002, 2003). The upflow shoots into the convectively stable photosphere, and is turned horizontally by the magnetic curvature forces along the tube. The dominating forces here are the centrifugal force of the flow, $\kappa \rho v^2$ and the magnetic curvature force, $\kappa B^2 / (4\pi)$, with $\kappa$ being the curvature. In equilibrium $v$ equals $v_{\rm A}$, with $v_{\rm A}$ being the Alfv{\'e}n velocity. During the evolution of the tube the velocity is roughly constant, but the magnetic field strength and hence $v_{\rm A}$ decreases leading to an overshoot, which creates an oscillation of the outflow around its equilibrium position such that the tube adopts a wave-like shape, i.e. the plasma first shoots up and then down, again passing the equilibrium position. Such a wave can be considered quasi-stationary, and the crest of such a wave can be compared with the properties of a Siphon flow (see below).
Hence, the flow yields a serpentine shape, looking like a sea serpent, and evidence for such radially aligned up and down flows has been presented by Sainz Dalda \& Bellot Rubio (2008).
The amplitude of this wave increases as the magnetic field strength decreases, and eventually the downflow part dives in the sub-photosphere. There the stratification is convectively unstable and the magnetic flux tube experiences a dynamic evolution, that produces outward propagating waves. This scenario produces down-flows and makes the tube to disappear within the penumbra. Thereby it would solve a problem of the moving tube model: the out-flow would not extend into the surrounding canopy, but would disappear within the penumbra, as it is observed.

\paragraph{Serpentine flow:} Such a two-dimensional serpentine solution was criticized to be unstable in three dimensions (Thomas 2005), arguing that buoyancy forces make the wavy tube to fall over sideways. But this argument is not valid, since the influence of the upflow at the footpoint of the tube is not taken into account. At the footpoint the plasma is ejected upwards into the photosphere and due to conservation of momentum, the plasma overshoots and follows an up- and down wavy behavior. The fact that the density at the upper crest is larger than in the surroundings does not make the tube to fall over. As an analogy, one may think of a jet of water directed upwards with a garden hose. As long as the jet is pointing upwards with the hose (at the footpoint), the jet of water will not fall over. The jet of water will not fall over, even though the density of water is larger than that of the surrounding air. Since the footpoint of the up-flowing flux tube and  its inclination is constrained, the boundary condition circumvents the wavy flow to fall over. Therefore the argument of Thomas (2005) is only true for a serpentine flow without a footpoint, and is not applicable here.

\paragraph{Siphon flows:} Siphon flow arches are stationary magnetic flux tube models, which were proposed to explain the Evershed flow (e.g., Meyer \& Schmidt 1968, Thomas 1988, Degenhardt 1991, Thomas \& Montesinos 1991). This class of models makes the ad hoc assumption of different magnetic field strengths at the two foot-points of a magnetic arch, which is responsible for a gas pressure gradient along the tube driving the flow. In the dynamic sea-serpent solutions (see above) a quasi-stationary solution exists (Schlichenmaier 2003). This solution corresponds to one (out of four) particular Siphon flow solution: a flow with a supercritical flow speed along the arch.

\paragraph{Heat transport:} Temporal measurements of the intensity evolution rule out the existence of interchange convection (Solanki \& R\"uedi 2003), and also, the numerical work of the moving tube model did not confirm the concept of interchange convection of magnetic flux tubes as the heating mechanism for the surplus brightness of the penumbra: A crucial result of the numerical investigation is that a tube rises and develops an upflow, but the upflow does not stop nor does the tube sink back down to the magnetopause. Hence, instead of interchange convection the moving tube simulations suggests that the heating occurs in form of upflow channels along magnetic field lines. Ruiz Cobo \& Bellot Rubio (2008) demonstrate that such an up-flow is capable to account for the brightness of the penumbra and that such up-flows can produce dark-cored bright filaments with a length of up to 3 Mm. Yet, even if such up flows can transport enough heat to account for the brightness of the penumbra, Schlichenmaier \& Solanki (2003) have shown, that downflows within the penumbra are obligatory: There is not enough space for the magnetic flux associated with the up flows, such that down flows must remove the magnetic flux from the photosphere. In this respect, the overshoot scenario (serpentine flow) may help: the hot up-flow cools and the cool down-flow heats up in the hot sub-photosphere, and re-enters the photosphere as a hot upflow. Hence, the moving tube scenario encounters problems in accounting for sufficient heat transport, but there are ways to solve the heat transport problem with channeled flows. And these channeled flows are driven by radiative cooling.

\subsection{Non-ideal magneto-hydrodynamics}

Before we continue with well accepted model descriptions of the penumbra, we also want to mention an off-track approach by Kuhn \& Morgan (2006) who argue that in the photosphere of a cool spot a large fraction of the plasma is neutral. A simplified consideration with two fluid components, one neutral and one ionized, yields an outward plasma flow driven by osmotic pressure. Whether or not such an effect and such a non-magnetic flow is realized on the Sun is not known, but it does not explain the observed Evershed flow, since the observed flow is magnetized.

\subsection{Radiative magneto-convection}

A better understanding of the sunspot penumbra is expected from numerical simulations of radiative magneto-convection in inclined magnetic fields. First results of such simulations (Heinemann et al. 2007; Scharmer et al. 2008, Rempel, Sch\"ussler, \& Kn\"olker 2008) consider 3D boxes solving for the full set of MHD equations including the (grey) radiative transport. These simulations consider a slice through the diameter of a round sunspot, including the umbra, the penumbra, and the surrounding quiet Sun. Assuming that the penumbral filament width is very small relative to the radius, the slice has a rectangular geometry with periodic boundary conditions in the horizontal directions. These simulations are still not able to produce a mature penumbra, but they succeed in reproducing single elongated filaments with lengths of up to a few Mm which resemble in many ways what is observed as thin light bridges and penumbral filaments of the inner penumbra.

The heating of these filaments does not occur by a single hot upflow channel, but rather in a form of a vertically elongated convective roll: a central lane of upflow, associated with two adjacent lanes of donwnflow. One convective cell has a vertical extension of some 500 km, while its lateral thickness is little less than 500 km. The vertical component of the magnetic field seems to become expelled by the convective flow, such that the convective cell is associated with more horizontal magnetic field. In this sense, these simulation results are a revival of the convective rolls proposed by Danielson in 1961. But in contrast to the Danielson rolls, the filaments in the simulations are interlaced with less inclined stronger magnetic field than in the filaments, and the rolls are elongated in depth.

The simulated penumbral filaments resemble light bridges with an inner upflow and two lateral downflows forming two apparent rolls as observed e.g. by Rimmele (2008). As the magnetic field becomes more inclined relative to the vertical, the upflow has an increasing horizontal outflow component. This horizontal outflow component increases with heigth, culminating in the photosphere. This horizontal flow is already present in the umbral dot simulations of Sch\"ussler \& V\"ogler (2006), but in an environment of more inclined magnetic fields this horizontal flow component becomes stronger. However, at this stage, the horizontal velocities in the simulations are only a little larger than the vertical velocities of the convective roll, while the state-of-the-art observations retrieve a horizontal velocity that is roughly a factor of 20 (10\,km/s compared to  $<0.5$\,km/s) larger than the vertical velocity.  In that sense the simulation fail to reproduce the Evershed effect, but there are indications that the simulated horizontal velocity component increases with more inclined magnetic field, and future simulations that may exhibit a fully developed penumbra and more inclined magnetic fields are expected to develop stronger horizontal flows, thereby reproducing the Evershed flow.

In the present MHD simulations the energy transport in umbral dots, light bridges, and filaments in the inner penumbra is accomplished by a magneto-convective mode, which may be characterized as convective elongated cells. Yet, these simulations do not exhibit a mature penumbra and the associated Evershed flow. It remains to be seen whether this magnetoconvective mode is also capable to reproduce a mature penumbra, or wether another magnetoconvective mode exists in the outer penumbra.

\section{Conclusions}

To explain different aspects of the penumbral properties two "simple" model classes have been proposed for the penumbra: (a) The moving magnetic flux tube models assumes ideal MHD, in which flows channeled by magnetic fields account for the filamentation, the Evershed flow, and the line asymmetries. (b) The gappy penumbra and convective rolls, which assume elongated convection cells to account for the surplus brightness of the penumbra. Neither of these simple models can account for all observational aspects: The moving tube scenario has problems to reproduce the overall down-flow in the outer penumbra and to account for all of the energy transport, while the elongated convective cells fail to produce an Evershed flow. The recent 3D box simulations of the full set of MHD equations show that magneto-convective heat transport may take place in elongated pancakes similar to what was proposed by Danielson. These simulations produce elongated convective cells which are associated with horizontal magnetic fields with weaker strengths, but in contrast to Danielson's proposal the horizontal rolls with horizontal magnetic fields are embedded in stronger and more vertical magnetic field. This latter magnetic field component would form a more or less static background.

Considering the state-of-the-art of observational results and theoretical modeling, we conclude that there is evidence for both, a channeled flow with velocities of more than 5\,km/s producing the Evershed flow, while the new 3D simulations suggest the existence of elongated convective rolls with much smaller up- and downflow velocities in the order 0.5 km/s. 
It may well be that the penumbra is a superposition of both, channeled flows above convective rolls, but at this point, we do not know.
More advanced simulations will ultimately produce a fully developed penumbra, and by then it will be possible to understand how an ensemble of dynamic filaments is capable to form a stable penumbra, and how the penumbra is heated. Finally, the questions of sunspot stability and how sunspots form, evolve, and decay can be addressed.

Yet, at this point, the models are not fully consistent with obervational facts. In particular it remains to be seen whether the flow pattern of convective rolls can be measured, and whether the  observed penumbral line asymmetries in the Stokes parameters including the NCP can be reproduced by such models. Spectropolarimetric measurements need to have a spatial resolution of better than 0.1 arcsec to be comparable to the models.

\begin{acknowledgements}
I am greatful to Oskar Steiner for many fruitful discussions, I like to thank Morten Franz for preparing Fig.~1, and Luis Bellot Rubio and Wolfgang Schmidt for valuable comments on the manuscript. 
\end{acknowledgements}


\def\apj{ApJ } 
\def\apjl{ApJL } 
\def\aap{A\&A } 
\def\nat{Nature }
\def\mnras{MNRAS }
\def\pasj{PASJ } 
\def\solphys{Solar Phys. }

\newcommand{\bysame}{\_\_\_\_\_}
\newcommand{\etalchar}[1]{$#1$}

\renewcommand{\bibitem}[1]{\item}
\begin{list}{}{\itemindent -1em \labelwidth 0em \labelsep 0em\leftmargin 1em}

\bibitem{2008A&A...480..825B}%
C.~{Beck}, \emph{{A 3D sunspot model derived from an inversion of
  spectropolarimetric observations and its implications for the penumbral
  heating}}, \aap \textbf{480} (2008), 825--838.

\bibitem{bellot2003}
L.~R. {Bellot Rubio}, \emph{The Fine Structure of the Penumbra: from Observations to Realistic Physical Models}, ASP conference series \textbf{307} (2003), 301.

\bibitem{bellot2004}%
L.~R. {Bellot Rubio}, \emph{{Sunspots as seen in polarized light}}, Rev. of
  Mod. Astron. \textbf{17} (2004), 21--50.

\bibitem{bellot+etal2003}%
L.~R. {Bellot Rubio}, H.~{Balthasar}, M.~{Collados}, and R.~{Schlichenmaier},
  \emph{{Field-aligned Evershed flows in the photosphere of a sunspot
  penumbra}}, \aap \textbf{403} (2003), L47--L50.

\bibitem{bellot+balthasar+collados2004}%
L.~R. {Bellot Rubio}, H.~{Balthasar}, and M.~{Collados}, \emph{{Two magnetic
  components in sunspot penumbrae}}, \aap \textbf{427} (2004), 319--334.

\bibitem{bellot+langhans+schlichenmaier2005}%
L.~R. {Bellot Rubio}, K.~{Langhans}, and R.~{Schlichenmaier}, \emph{{Multi-line
  spectroscopy of dark-cored penumbral filaments}}, \aap \textbf{443} (2005),
  L7--L10.

\bibitem{bellot+schlichenmaier+tritschler2006}%
L.~R. {Bellot Rubio}, R.~{Schlichenmaier}, and A.~{Tritschler},
  \emph{{Two-dimensional spectroscopy of a sunspot. III. Thermal and kinematic
  structure of the penumbra at 0.5 arcsec resolution}}, \aap \textbf{453}
  (2006), 1117--1127.
  
\bibitem{bellot+etal2007}%
L.~R.~Bellot Rubio, S.~Tsuneta, K.~Ichimoto, Y.~Katsukawa, B.~W.~Lites, S.~Nagata, T.~Shimizu, R.~A.~Shine, Y.~Suematsu, T.~D.~Tarbell, A.~M.~Title, and J.~C.~del Toro Iniesta, 
\emph{Vector spectropolarimetry of dark-cored penumbral filaments with Hinode},
\apjl \textbf{668} (2007), L91--L94.

\bibitem{2007ApJ...669L..57B}%
L.~{Bharti}, C.~{Joshi}, and S.~N.~A. {Jaaffrey}, \emph{{Observations of Dark
  Lanes in Umbral Fine Structure from the Hinode Solar Optical Telescope:
  Evidence for Magnetoconvection}}, \apjl \textbf{669} (2007), L57--L60.

\bibitem{borrero+etal2004}%
J.~M. {Borrero}, S.~K. {Solanki}, L.~R. {Bellot Rubio}, A.~{Lagg}, and S.~K.
  {Mathew}, \emph{{On the fine structure of sunspot penumbrae. I. A
  quantitative comparison of two semiempirical models with implications for the
  Evershed effect}}, \aap \textbf{422} (2004), 1093--1104.

\bibitem{borrero+etal2005}%
J.~M. {Borrero}, A.~{Lagg}, S.~K. {Solanki}, and M.~{Collados}, \emph{{On the
  fine structure of sunspot penumbrae. II. The nature of the Evershed flow}},
  \aap \textbf{436} (2005), 333--345.

\bibitem{borrero+etal2006}%
J.~M. {Borrero}, S.~K. {Solanki}, A.~{Lagg}, H.~{Socas-Navarro}, and
  B.~{Lites}, \emph{{On the fine structure of sunspot penumbrae. III. The
  vertical extension of penumbral filaments}}, \aap \textbf{450} (2006),
  383--393.

\bibitem{borrero+etal2007}%
J.~M.~Borrero, L.~R.~Bellot Rubio, and D.~A.~N.~M\"uller, 
\emph{Flux Tubes as the Origin of Net Circular Polarization in Sunspot Penumbrae},
 \apj \textbf{666} (2007), L133--L136.

\bibitem{borrero+etal2008}%
J.~M. {Borrero}, B.~W.~Lites, and S.~K. {Solanki}, \emph{{Evidence of magnetic field wrapping around penumbral filaments}}, \aap \textbf{481} (2008), L13-L16.

\bibitem{borrero+solanki2008}
J.~M. {Borrero} and S.~K. {Solanki}, \emph{Are there field-free gaps near tau=1 in sunspot penumbrae?}, \apj in press (2008).

\bibitem{cabrera+etal2008}
D.~Cabrera Solana, L.~R.~Bellot Rubio, J.~M.~Borrero, J.-C.~del Toro Iniesta, \emph{Temporal evolution of the Evershed flow in sunspots. II. Physical properties and nature of Evershed clouds}, \aap \textbf{477} (2008), 273--283.

\bibitem{danielson1961}%
R.~E. {Danielson}, \emph{{The Structure of Sunspot Penumbras. II.
  Theoretical.}}, \apj \textbf{134} (1961), 289.

\bibitem{degenhardt1991}%
D.~{Degenhardt}, \emph{Stationary fiphon flows in thin magnetic flux tubes.
  ii}, \aap \textbf{248} (1991), 637.

\bibitem{deinzer1965}%
W.~{Deinzer}, \emph{{On the Magneto-Hydrostatic Theory of Sunspots.}}, \apj
  \textbf{141} (1965), 548--563.

\bibitem{delToro+etal2001}
J.~C.~del Toro Iniesta, L.~R.~ Bellot Rubio, M.~ Collados,
\emph{Cold, Supersonic Evershed Downflows in a Sunspot},
\apjl \textbf{549} (2001), L139-L142

\bibitem{evershed1909}%
J.~{Evershed}, \emph{Radial movement in sun-spots}, \mnras \textbf{69} (1909),
  454--457.

\bibitem{grosser1991}%
H.~{Grosser},
\emph{Zur Entstehung der Penumbra-Filamentierung von Sonnenflecken durch die Wirkung von Konvektionsrollen}, 
PhD thesis, Universit\"at G\"ottingen (1991). 

\bibitem{heinemann+etal2007}%
T.~{Heinemann}, {\AA}.~{Nordlund}, G.~B. {Scharmer}, and H.~C. {Spruit},
  \emph{{MHD Simulations of Penumbra Fine Structure}}, 
  \apj \textbf{669}
  (2007), 1390--1394.

\bibitem{ichimoto+etal2007a}%
K.~{Ichimoto}, R.~A. {Shine}, B.~{Lites}, M.~{Kubo}, T.~{Shimizu},
  Y.~{Suematsu}, S.~{Tsuneta}, Y.~{Katsukawa}, T.~D. {Tarbell}, A.~M. {Title},
  S.~{Nagata}, T.~{Yokoyama}, and M.~{Shimojo}, \emph{{Fine-Scale Structures of
  the Evershed Effect Observed by the Solar Optical Telescope aboard Hinode}},
  \pasj \textbf{59} (2007a), 593.

\bibitem{ichimoto+etal2007b}%
K.~{Ichimoto}, Y.~{Suematsu}, S.~{Tsuneta}, Y.~{Katsukawa}, T.~{Shimizu}, R.~A.
  {Shine}, T.~D. {Tarbell}, A.~M. {Title}, B.~W. {Lites}, M.~{Kubo}, and
  S.~{Nagata}, \emph{{Twisting Motions of Sunspot Penumbral Filaments}},
  Science \textbf{318} (2007b), 1597.

\bibitem{ichimoto+etal2008}%
K.~{Ichimoto}, S.~{Tsuneta}, Y.~{Suematsu}, Y.~{Katsukawa}, T.~{Shimizu}, B.~W.
  {Lites}, M.~{Kubo}, T.~D. {Tarbell}, R.~A. {Shine}, A.~M. {Title}, and
  S.~{Nagata}, \emph{{Net circular polarization of sunspots in high spatial
  resolution}}, \aap \textbf{481} (2008), L9--L12.

\bibitem{jahn1989}%
K.~{Jahn}, \emph{{Current sheet as a diagnostic for the subphotospheric
  structure of a SPOT}}, \aap \textbf{222} (1989), 264--292.

\bibitem{jahn+schmidt1994}%
K.~{Jahn} and H.~U. {Schmidt}, 
\emph{Thick penumbra in a magnetostatic sunspot model}, 
\aap \textbf{290} (1994), 295--317.
  
\bibitem{jahn1997}%
K.~{Jahn}, \emph{{Physical Models of Sunspots}}, ASP Conf. Ser. 118: 1st Advances
  in Solar Physics Euroconference. Advances in Physics of Sunspots
  (B.~{Schmieder}, J.~C. {del Toro Iniesta}, and M.~{Vazquez}, eds.), 1997,
  p.~122.

\bibitem{jurcak+etal2007}%
J.~Jurcak. L.~R.~Bellot Rubio,
K.~{Ichimoto}, Y.~{Katsukawa}, B.~W. {Lites},  S.~{Nagata}, T.~{Shimizu}, Y.~{Suematsu}, 
T.~D. {Tarbell}, A.~M. {Title}, and S.~{Tsuneta},
\emph{The analysis of penumbral fine structure using an advanced inversion technique},
PASJ \textbf{59} (2007), S601--S606.

\bibitem{jurcak+bellot2008}%
J.~Jurcak and L.~R.~Bellot Rubio,
\emph{Penumbral models in the light of Hinode spectropolarimetric 
observations}
\aap \textbf{481} (2008), L17--L20.

\bibitem{2006ASPC..354..230K}%
J.~R. {Kuhn} and H.~{Morgan}, \emph{{Osmotically Driven Neutral Sunspot
  Winds}}, Solar MHD Theory and Observations: A High Spatial Resolution
  Perspective (J.~{Leibacher}, R.~F. {Stein}, and H.~{Uitenbroek}, eds.),
  Astronomical Society of the Pacific Conference Series, vol. 354, December
  2006, pp.~230--+.

\bibitem{landolfi+landi1996}%
M.~{Landolfi} and E.~{Landi degl'Innocenti}, \emph{Net Circular Polarization
  in Magnetic Spectral Lines Produced by Velocity Gradients: Some Analytical
  Results}, \solphys \textbf{164} (1996), 191--202.

\bibitem{langhans+etal2005}%
K.~Langhans, G.~B.~Scharmer, D.~Kiselman, M.~G.~L\"ofdahl, and T.~E.~Berger, 
\emph{Inclination of magnetic fields and flows in sunspot penumbrae},
\aap \textbf{436} (2005), 1087--1101.

\bibitem{langhans+etal2007}%
K.~Langhans, G.~B.~Scharmer, D.~Kiselman, and M.~G.~L\"ofdahl, 
\emph{Observations of dark-cored filaments in sunspot penumbrae},
\aap \textbf{464} (2007), 763--774.

\bibitem{martinez2000}%
V.~{Mart{\'\i}nez Pillet}, \emph{Spectral signature of uncombed penumbral
  magnetic fields}, \aap \textbf{361} (2000), 734--742.

\bibitem{mueller+etal2002}%
D.~A.~N.~{M{\"u}ller}, R.~{Schlichenmaier}, O.~{Steiner}, and M.~{Stix},
  \emph{{Spectral signatures of magnetic flux tubes in sunspot penumbrae}},
  \aap \textbf{393} (2002), 305--319.
  
\bibitem{mueller+etal2006}%
D.~A.~N.~{M{\"u}ller}, R.~{Schlichenmaier}, G.~{Fritz}, and C.~{Beck},
  \emph{{The multi-component field topology of sunspot penumbrae. A diagnostic
  tool for spectropolarimetric measurements}}, \aap \textbf{460} (2006),
  925--933.

\bibitem{maltby1964}%
P.~{Maltby}, \emph{On the velocity field in sunspots}, Astrophysica Norvegica
  \textbf{8} (1964), 205.

\bibitem{meyer+schmidt1968}%
F.~{Meyer} and H.~U.~{Schmidt}, \emph{{Magnetisch ausgerichtete Str\"omungen
  zwischen Sonnenflecken}}, {Zeitschrift f\"ur angewandte Mathematik und
  Mechanik} \textbf{48} (1968), T218.

\bibitem{meyer+schmidt+weiss1977}%
F.~Meyer, N.~O.~Weiss, H.~U.~Schmidt, \emph{{The stability of sunspots}}, {MNRS} \textbf{179}
  (1977), 741.

\bibitem{1979ApJ...234..333P}%
E.~N.~{Parker}, 
\emph{{Sunspots and the physics of magnetic flux tubes. IX Umbral dots and longitudinal overstability}}, 
\apj \textbf{234} (1979), 333--347.

\bibitem{pizzo1990}%
V.~J.~{Pizzo}, 
\emph{Numerical modeling of solar magnetostatic structures bounded by current sheets}, 
\apj \textbf{365} (1990), 764--777.

\bibitem{rempel+etal2008}%
M.~Rempel, M.~Sch\"ussler, \& M.~Kn\"olker, 
\emph{Radiative MHD simulation of sunspot structure}, 
submitted to ApJ (2008).

\bibitem{rezaei+etal2006}
R.~Rezaei, R.~Schlichenmaier, C.~Beck, and L.~R.~Bellot Rubio,
\emph{The flow field in the sunspot canopy},
\aap \textbf{454} (2006), 975--982.

\bibitem{2004ApJ...604..906R}%
T.~{Rimmele}, 
\emph{{Plasma Flows Observed in Magnetic Flux Concentrations
and Sunspot Fine Structure Using Adaptive Optics}}, 
\apj \textbf{604} (2004), 906--923.

\bibitem{rimmele+marino2006}%
T.~{Rimmele} and J.~{Marino}, 
\emph{{The Evershed Flow: Flow Geometry and Its
  Temporal Evolution}}, \apj \textbf{646} (2006), 593--604.

\bibitem{2008ApJ...672..684R}%
T.~{Rimmele}, \emph{{On the Relation between Umbral Dots, Dark-cored Filaments,
  and Light Bridges}}, \apj \textbf{672} (2008), 684--695.

\bibitem{ruiz+bellot2008}
B.~Ruiz Cobo, L.~R.~Bellot Rubio, \emph{Heat transfer in sunspot penumbrae. Origin of dark-cored penumbral filaments}, \aap \textbf{488} (2008), 749--756.

\bibitem{2008A&A...481L..21S}%
A.~{Sainz Dalda} and L.~R. {Bellot Rubio}, \emph{{Detection of sea-serpent
  field lines in sunspot penumbrae}}, \aap \textbf{481} (2008), L21--L24.

\bibitem{sanchez+lites1992}%
J.~Sanchez Almeida, B.~W.~Lites, 
\emph{Observation and Interpretation of the asymmetric Stokes Q, U, and V line profiles in sunspots}, 
\apj \textbf{398} (1992), 359--374.

\bibitem{scharmer+spruit2006}%
G.~B. {Scharmer} and H.~C. {Spruit}, 
\emph{Magnetostatic penumbra models with field-free gaps}, 
\aap \textbf{460} (2006), 605--615.

\bibitem{scharmer+etal2002}%
G.~B. Scharmer, B.~V. Gudiksen, D.~Kiselman, M.~G. L{\"o}fdahl, and L.~H.~M. Rouppe van~der Voort, 
\emph{{Dark cores in sunspot penumbral filaments}},
\nat \textbf{420} (2002), 151.

\bibitem{scharmer+etal2008}%
G.~B. {Scharmer},  T.~{Heinemann}, and A.~{Nordlund}, 
\emph{Convection and the origin of Evershed flows in sunspot penumbrae}
\apj \textbf{677} (2008), L149--L152.

\bibitem{schlichenmaier+jahn+schmidt1998a}%
R.~{Schlichenmaier}, K.~{Jahn}, and H.~U. {Schmidt},
\emph{A dynamical model for the penumbral fine structure and the evershed effect in sunspots}, 
\apjl \textbf{493} (1998), L121.

\bibitem{schlichenmaier+jahn+schmidt1998b}%
R.~{Schlichenmaier}, K.~{Jahn}, and H.~U. {Schmidt}, 
\emph{Magnetic flux tubes evolving in sunspots. 
A model for the penumbral fine structure and the evershed flow}, 
\aap \textbf{337} (1998), 897--910.
  
\bibitem{schlichenmaier+schmidt2000}%
R.~{Schlichenmaier} and W.~{Schmidt}, 
\emph{Flow geometry in a sunspot penumbra}, 
\aap \textbf{358} (2000), 1122--1132.

\bibitem{schlichenmaier2002}%
R.~{Schlichenmaier}, 
\emph{{Penumbral fine structure: Theoretical understanding}}, 
AN \textbf{323} (2002), 303--308.

\bibitem{schlichenmaier+collados2002}%
R.~{Schlichenmaier} and M.~{Collados}, \emph{Spectropolarimetry in a sunspot
  penumbra. Spatial dependence of stokes asymmetries in fe I 1564.8 nm}, \aap
  \textbf{381} (2002), 668--682.

\bibitem{schlichenmaier2003p}%
R.~{Schlichenmaier}, \emph{The sunspot penumbra: new developments}, ASP Conf. Ser. 286:
  Current theoretical models and future high resolution solar observations
  (A.~{Pevtsov} and H.~{Uitenbroek}, eds.), 2003, pp.~211--226.

\bibitem{schlichenmaier+solanki2003}%
R.~{Schlichenmaier} and S.~K. {Solanki}, 
\emph{{On the heat transport in a sunspot penumbra}}, 
\aap \textbf{411} (2003), 257--262.

\bibitem{schlichenmaier+bellot+tritschler2004}%
R.~{Schlichenmaier}, L.~R. {Bellot Rubio}, and A.~{Tritschler},
  \emph{{Two-dimensional spectroscopy of a sunspot. II. Penumbral line
  asymmetries}}, \aap \textbf{415} (2004), 731--737.
 
\bibitem{schmidt+schlichenmaier2000}%
W.~{Schmidt} and R.~{Schlichenmaier}, \emph{Small-scale flow field in a sunspot
  penumbra}, \aap \textbf{364} (2000), 829--834.

\bibitem{2005A&A...441..337S}%
M.~{Sch{\"u}ssler} and M.~{Rempel}, \emph{{The dynamical disconnection of
  sunspots from their magnetic roots}}, \aap \textbf{441} (2005), 337--346.

\bibitem{2006ApJ...641L..73S}%
M.~{Sch{\"u}ssler} and A.~{V{\"o}gler}, \emph{{Magnetoconvection in a Sunspot
  Umbra}}, \apjl \textbf{641} (2006), L73--L76.

\bibitem{sobotka+suettlerin2001}
M.~{Sobotka} and P.~{S\"utterlin}, \emph{Fine structure in sunspots. IV. Penumbral grains in speckle reconstructed images}, \aap \textbf{380} (2001), 714--715.

\bibitem{socas+etal2004}
H.~Socas-Navarro, V.~Martinez Pillet, M.~Sobotka, M.~Vazquez, 
\emph{The thermal and magnetic structure of umbral dots form the inversion of high-resolution full Stokes Observations},
\apj \textbf{614}, (2004), 448--456.

\bibitem{solanki+montavon1993}%
S.~K. {Solanki} and C.~A.~P. {Montavon}, \emph{Uncombed fields as the source of
  the broad-band circular polarization of sunspots}, \aap \textbf{275} (1993),
  283.

\bibitem{solanki+rueedi+livingston1992a}%
S.~K. {Solanki}, I.~{Rueedi}, and W.~{Livingston}, \emph{{Infrared lines as
  probes of solar magnetic features. V - The magnetic structure of a simple
  sunspot and its canopy}}, \aap \textbf{263} (1992), 312.

\bibitem{2003A&A...411..249S}%
S.~K. {Solanki} and I.~{R{\"u}edi}, \emph{{Spatial and temporal fluctuations in
  sunspots derived from MDI data}}, \aap \textbf{411} (2003), 249--256.

\bibitem{solanki2003}%
S.~K. {Solanki},
\emph{Sunspots: An overview},
The Astronomy and Astrophysics Review \textbf {11} (2003),153--286.

\bibitem{spruit+scharmer2006}%
H.~C. {Spruit} and G.~B. {Scharmer}, \emph{{Fine structure, magnetic field and
  heating of sunspot penumbrae}}, \aap \textbf{447} (2006), 343--354.

\bibitem{2000SoPh..196..245S}%
O.~{Steiner}, \emph{{The formation of asymmetric Stokes V profiles in the
  presence of a magnetopause}}, \solphys \textbf{196} (2000), 245--268.

\bibitem{suetterlin+etal2004}%
P.~S\"utterlin, L.~R.~Bellot Rubio, and R.~Schlichenmaier,
\emph{Asymmetrical appearance of dark-cored filaments in sunspot penumbrae},
\aap \textbf{424} (2004), 1049-1053.

\bibitem{thomas1988}%
J.~H. {Thomas},
\emph{Siphon flows in isolated magnetic flux tubes},
\apj \textbf{333} (1988), 407--419.
  
\bibitem{1991ApJ...375..404T}%
J.~H. {Thomas} and B.~{Montesinos}, 
\emph{{Siphon flows in isolated magnetic flux tubes. IV Critical flows with standing tube shocks}}, 
\apj \textbf{375} (1991), 404--413.

\bibitem{2005A&A...440L..29T}%
J.~H. {Thomas}, \emph{{Flows along penumbral flux tubes in sunspots.
  Instability of super-Alfv{\'e}nic, serpentine solutions}}, \aap \textbf{440}
  (2005), L29--L32.

\bibitem{title+etal1993}
A.~M.~Title,  Z.~A~Frank, R.~A.~Shine, T.~D.~Tarbell, K.~P.~Topka, G.~Scharmer, W.~Schmidt,
\emph{On the magnetic and velocity field geometry of simple sunspots},
\apj \textbf{403} (1993), 780--796.

\bibitem{tritschler+etal2004}%
A.~{Tritschler}, R.~{Schlichenmaier}, L.~R. {Bellot Rubio}, and {the KAOS
  Team}, \emph{{Two-dimensional spectroscopy of a sunspot. I. Properties of the
  penumbral fine structure}}, \aap \textbf{415} (2004), 717--729.

\bibitem{tritschler+etal2007}%
A.~{Tritschler}, D.~A.~N. {M{\"u}ller}, R.~{Schlichenmaier}, and H.~J.
  {Hagenaar}, \emph{{Fine Structure of the Net Circular Polarization in a
  Sunspot Penumbra}}, \apjl \textbf{671} (2007), L85--L88.

\bibitem{wiehr1995}%
E.~ Wiehr, 
\emph{The origin of the Evershed asymmetry}, 
\aap \textbf{298} (1995), L17--L20.

\bibitem{weiss1964}%
N.~O.~Weiss, 
\emph{Magnetic flux tubes and convection in the Sun}, 
{MNRS} \textbf{128} (1964), 225.

\bibitem{westendorp+etal2001a}%
C.~{Westendorp Plaza}, J.~C. {del Toro Iniesta}, B.~{Ruiz Cobo}, and
  V.~{Mart{\'\i}nez Pillet}, \emph{{Optical Tomography of a Sunspot. III.
  Velocity Stratification and the Evershed Effect}}, \apj \textbf{547} (2001),
  1148--1158.

\bibitem{westendorp+etal2001b}%
C.~{Westendorp Plaza}, J.~C. {del Toro Iniesta}, B.~{Ruiz Cobo},
  V.~{Mart{\'\i}nez Pillet}, B.~W. {Lites}, and A.~{Skumanich}, \emph{{Optical
  Tomography of a Sunspot. II. Vector Magnetic Field and Temperature
  Stratification}}, \apj \textbf{547} (2001), 1130--1147.

\bibitem{zakharov+etal2008}%
V.~Zakharov, J.~Hirzberger, T.~L.~Riethm\"uller, S.~K.~Solanki, and P.~Kobel, \emph{Evidence of convective rolls in a sunspot penumbra}, (2008), submitted to \aap, arxiv:0808.2317v1.

\end{list}

\end{document}